\documentclass[runningheads, 10pt]{llncs}
\usepackage{graphicx}
\usepackage[letterpaper, margin=1.0in]{geometry}

\usepackage{hyperref}
\usepackage{xcolor}
\usepackage{amsmath}
\usepackage{amsfonts}
\usepackage{amssymb}
\usepackage{subfigure}
\usepackage{nicefrac}
\usepackage{multirow, tabularx, booktabs}
\usepackage{comment}
\usepackage{amsfonts}
\usepackage{xspace}
\usepackage{listings}
\usepackage{xcolor}

\usepackage[citestyle=numeric,sorting=none,bibstyle=authoryear]{biblatex}
\AtEveryBibitem{[\printfield{labelnumber}]\addspace}

\newcommand{\rnaglib}{\texttt{rnaglib}\xspace}

\definecolor{codegreen}{rgb}{0,0.6,0}
\definecolor{codegray}{rgb}{0.5,0.5,0.5}
\definecolor{codepurple}{rgb}{0.58,0,0.82}
\definecolor{backcolour}{rgb}{0.95,0.95,0.92}

\lstdefinestyle{mystyle}{
    language=Python,
    basicstyle=\ttfamily\footnotesize,
    breakatwhitespace=false,         
    frame=single,
    breaklines=true,                 
    captionpos=b,                    
    keepspaces=true,                 
    showspaces=false,                
    showstringspaces=false,
    showtabs=false,                  
    tabsize=2
}

\begin{document}
%
\title{3D-based RNA function prediction tools in \rnaglib}

\titlerunning{3D RNA function prediction tools with rnaglib}
%
\author{Carlos Oliver\inst{1}
\and{Vincent Mallet} \inst{2}
\and J\'er\^ome Waldisp\"uhl \inst{3}
}
\authorrunning{Oliver et al.}
%
\institute{
1. Department of Machine Learning and Systems Biology, Max Planck Institute of Biochemistry, Martinsried, Germany\\
\email{oliver@biochem.mpg.de}\\
2. LIX, \'Ecole Polytechnique, Paris, France\\
3. School of Computer Science, McGill University, Montr\'eal, Canada \\
}

\maketitle              
\begin{abstract}

Understanding the connection between complex structural features of RNA and biological function is a fundamental challenge in evolutionary studies and in RNA design.
However, building datasets of RNA 3D structures and making appropriate modeling choices remains time-consuming and lacks standardization.
In this chapter, we describe the use of \rnaglib, to train supervised and unsupervised machine learning-based function prediction models on datasets of RNA 3D structures.


\end{abstract}
\newpage
\section{Introduction}

In the protein domain, which still enjoys a substantial advantage in number of solved structures over RNA, deep learning approaches have made significant advances in protein property prediction \cite{deepfri} and design \cite{baker}.  
Along with larger datasets, the methods developed for proteins were propelled by the development of sophisticated representations of structures which exposed them to geometric deep learning methods such as graph neural networks \cite{graphgen}, and large-scale self-supervised training \cite{esm}.

RNAs are distinct from proteins and have their own specificity. In particular, they fold hierarchically. A secondary structure made of Watson-Crick and Wobble base pairs is rapidly assembled to serve as a scaffold for 3D structures \cite{how}. During that process, non-canonical base interactions \cite{lw} are formed to stabilize the tertiary architecture. This information, along with others that will be discussed below, provides rich graphical models that can be used to develop customized solutions for analyzing and designing RNA structures.

Building on these customized solution and increasingly available RNA structural data, opportunities to uncover structure-function relationships and validate design algorithms through data-driven approaches for RNA come within reach \cite{rnamigos}. 
With \rnaglib \cite{rnaglib}, we take a step towards these advances by providing a powerful graph encoding of RNA 3D structures, self-supervised learning training regimes, and dataset construction utilities to lower the barrier to entry. 
 
\subsection{RNA 3D Data in \rnaglib}

In raw form, the RNA structure is a set of 3D coordinates for every atom in the molecule.
Typically obtained from X-ray crystallography.
Associated to the structural data, \rnaglib also collects external annotations such as small molecule binding partners, protein interfaces, secondary structure elements, chemical modifications, etc. 
These annotations or any others which can be collected by the user typically form the target space of a prediction task.
Currently we host a set of 5759 total structures and 1176 filtered for redundancy.

\subsection{Representing RNA 3D structures}

\begin{figure}
\centering
\includegraphics[width=.8\textwidth]{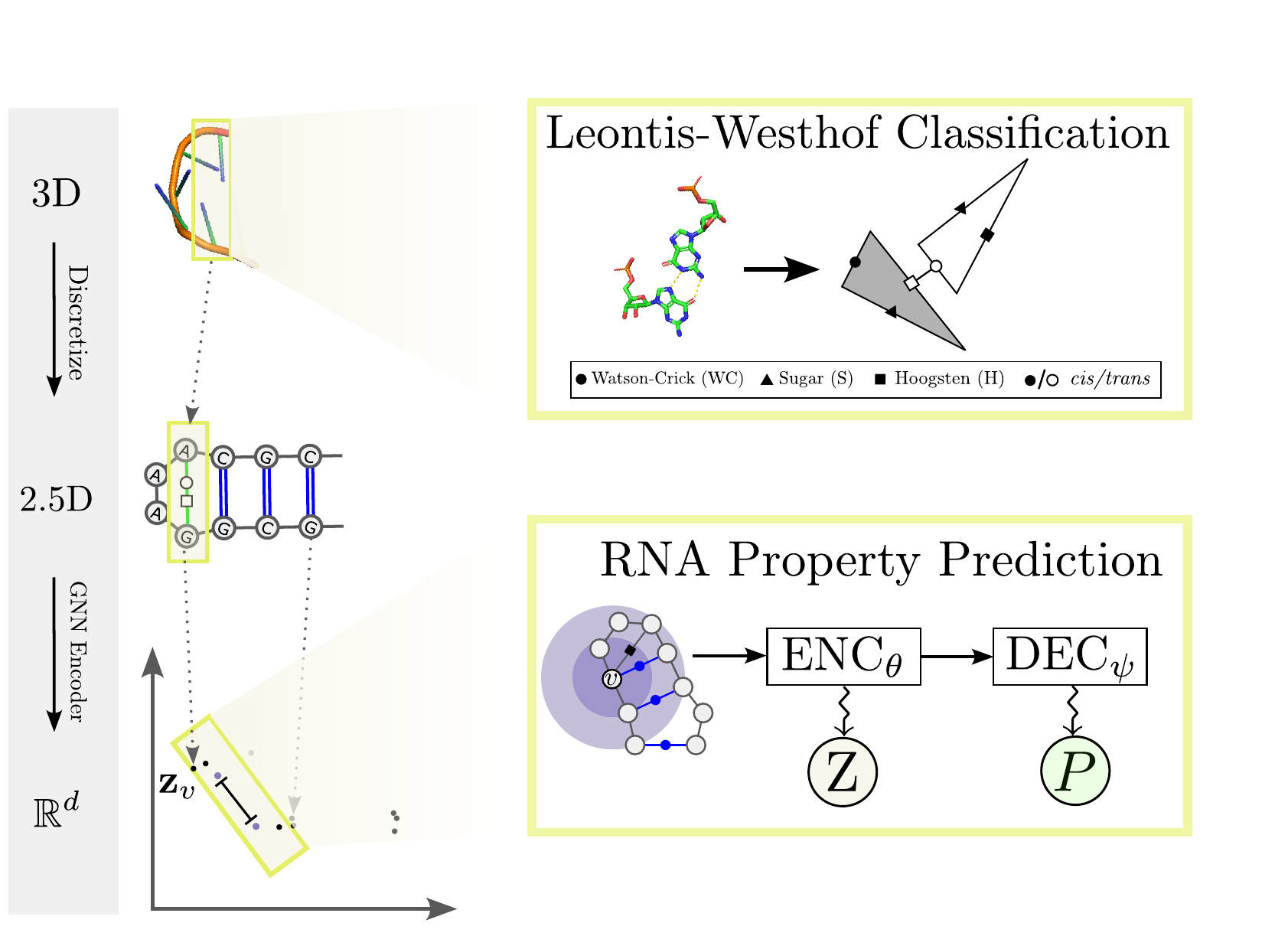}
\caption{RNA 3D representation learning paradigm. We use the Leontis-Westhof base pair geometry classification to build expressive graphs of RNA 3D models. The graphs are then embedded into a learned space with an encoder network $\mathrm{ENC}_\theta$ to emit embedding space $\mathbf{Z}$ and property space $\mathbf{P}$ with the decoder $\mathrm{DEC}_\psi$.}
\label{fig:lw}
\end{figure}

In order to work with deep learning methods, we need a \emph{biophysical model} compatible with state of the art data-driven methods.
Specifically for RNA, the set of basepair interactions naturally leads to a graph representation where each node is a base, and an edge is a basepair.
For ligand binding, it is known that detailed geometric orientations of binding sites is crucial and the typical secondary structure encoding can often lose this signal \cite{rnamigos}.
For this reason, a promising approach is to use the Leontis-Westhof classification which is made available through software packages such as \texttt{x3dna-dssr} \cite{dssr} and \texttt{FR3D} \cite{fr3d}.
See Fig. \ref{fig:lw} top panel for an overview of the representation learning workflow.

Having built an expressive encoding of the RNA 3D structure, that is here denoted as RNA 2.5D structure, we can leverage encoder-decoder representation learning methods \cite{ham} to predict functional attributes of interest (e.g. small-molecule binding, chemical modifications, protein binding, etc.) as shown in Figure \ref{fig:lw}.
Here, we  rely on an $\textrm{ENC}_\phi$ model parametrized by $\phi$, typically a neural network $\rightarrow \mathbb{R}^d$ which map nodes in a graph to real-valued embedding vectors $\mathbf{z}_{u}$ for each node $u \in G$.
At this stage, the encoder can already be trained through self-supervised paradigm such as those which enforce distances in the embedding space to reflect those of a domain-specific similarity function on the graph space.
For more details on this paradigm we point the reader to \cite{vernal} and note that advanced functionalities of \rnaglib provide support for self supervised learning but are not covered in this chapter.
Adding a trainable output layer ($\textrm{DEC}_\psi$) to the RNA embedding models which is tuned for a given dataset allows pre-trained models to perform downstream property prediction tasks.

\subsection{Objectives} 

The objective of this tutorial is to familiarize RNA informatics and machine learning practitioners with the functionalities of \rnaglib, which support RNA 3D dataset access and construction as well as machine learning model training and evaluation.

\section{Materials}

\rnaglib is an open source Python package hosted on GitHub at \url{https://github.com/cgoliver/rnaglib}.
Full package documentation is available at \url{https://rnaglib.readthedocs.io/en/latest/}.
\rnaglib can be installed through the PyPi command line utility as follows:

\begin{lstlisting}
$ pip install rnaglib
\end{lstlisting}

or the latest version of the library can be installed from source with:

\begin{lstlisting} 
$ git clone https://github.com/cgoliver/rnaglib 
$ pip install .
\end{lstlisting}

Optional dependencies consist mainly of deep learning frameworks which are left to the user to install to make the installation framework-agnostic.
We currently support model training with DGL and Pytorch Geometric.
Familiarity with Python, basic command-line usage, and common deep learning frameworks is assumed.

\section{Dataset Utilities}

In this tutorial we will load an RNA dataset and use it to train a self-supervised model with the downstream task of predicting small-molecule binding site residues.
Along the way, we will cover some useful features of \rnaglib and refer the user to the package documentation for full details.

\subsection{Dataset preparation}\label{sec:obtain}

\subsubsection{Obtaining annotated structures}

After installing \rnaglib, a command-line utility is made available to provide access to annotated datasets of RNA 3D structures hosted on Zenodo \footnote{\url{https://zenodo.org/records/7624873}}.

\begin{lstlisting}
$ rnaglib_download --redundancy all --version 1.0.0
\end{lstlisting}

All data is downloaded by default to \url{~/.rnaglib} unless specified otherwise with the \texttt{--download\_dir} flag.
By specifying \texttt{--redundancy [nr, all]}, you select whether to download a complete copy of all annotated 3D structures available at the time of building in the PDB or, with \texttt{--redundancy nr}, a representative set built by the BGSU RNA 3D Hub \cite{fr3d}.
The \texttt{--version} specifies which release of the datasets to use with higher version numbers reflecting more recent builds.

\subsubsection{Annotating your own structure}

If you have a local or custom PDB you would like to annotate for use with \rnaglib, we provide the function \texttt{fr3d\_to\_graph} which returns a NetworkX object containing Leontis-Westhof annotations \footnote{ Currently this step is done using fr3d-python \url{https://github.com/BGSU-RNA/fr3d-python} which is in beta.}.

\begin{lstlisting}
from rnaglib.prepare_data import fr3d_to_graph 
pdb_path = "../data/1aju.cif"
annots = fr3d_to_graph(pdb_path)
\end{lstlisting}

\subsection{Inspecting the RNA 3D object}

Each RNA is stored as a dictionary with each key mapping to some basic information about the molecule.
Upon creation, the dictionary contains the ID, and a networkx graph object of its interactions, and a path to the source data file.

\begin{lstlisting}
>>> from rnaglib.utils import available_pdbids
# returns a list of PDBIDs
>>> pdbids = available_pdbids()
# get the first RNA by PDBID
>>> graph_from_pdbid(pdbids[0])
{"rna_name": "1a9n.json",
 "rna": <networkx.classes.digraph.DiGraph object at 0x1567ae620>, 
 "path": "~/.rnaglib/datasets/rnaglib-nr-1.0.0/graphs/1a9n.json"}

\end{lstlisting}

The \texttt{"rna"} key holds all the annotations for the RNA as a NetworkX object.
Each node and edge in the object is keyed by a unique identifier in the format \texttt{"<pdbid>.<chain>.<residue number>"} (e.g. \texttt{"1a9n.Q.0"}), which can be used to access all nucleotide, base pair and RNA-level annotations.

\begin{lstlisting}
>>> data["rna"].edges[("1a9n.Q.0", "1a9n.Q.1")]     
{"LW": "B53", "backbone": True}
>>> data.nodes["1a9n.Q.0"]["nt_code"]
"C"
>>> data.graph
{'dbn': {'all_chains': {'num_nts': 28, 'num_chars': 29, 'bseq': 'AACGGGCGCAGAA&UCUGACGGUACGUUU', ... }
\end{lstlisting}

\subsection{The RNA 3D Dataset object}

To make dealing with more than a single RNA at a time more convenient, we provide the \texttt{RNADataset} object to hold collections of RNAs.

\begin{lstlisting}
>>> from rnaglib.data_loading import RNADataset
>>> dataset = RNADataset()
\end{lstlisting}

The above invocation by default returns the most recent non-redundant dataset available.
To specify a given version of the dataset, you can specify \texttt{"version"} and \texttt{"redundancy"} through keyword arguments.
Datasets can be indexed like a list or you can inspect an individual RNA by its PDBID.

\begin{lstlisting}
>>> rna_1 = dataset[3]
>>> pdbid = dataset.available_pdbids[3]
>>> rna_2 = dataset.get_pdbid(pdbid)
\end{lstlisting}

The RNA dataset can be filtered using any logic you like (through the \texttt{RNADataset.subset()} method), for example by resolution, year, secondary structure element, sequence identity, etc. to name a few, thereby creating a new dataset.

\section{Machine Learning Tutorial}

At this point we have a hold of collections of RNA structures and associated annotations in the \texttt{RNADataset} object.
Though, in order to perform function prediction, we need two more steps. First, building vectorial representations of the RNA structure and next, feeding the data to a traineable model.

\subsection{RNA Representations}

The \texttt{Representation} object has the job of generating an ML-friendly vectorial representation from the raw annotated RNA as well as any features and associated prediction targets.
Currently available representations are graphs, point clouds, and voxels.
At this point we need to specify which learning framework to use for building the representations.
The choice of framework should match that of the implementation of the model. We currently support PyTorch Geometric and DGL.

\begin{lstlisting}
>>> from rnaglib.representations import GraphRepresentation

>>> graph_rep = GraphRepresentation(framework="dgl")
>>> nt_features = ["nt_code"]
>>> nt_targets = ["binding_ion"]
>>> dataset = RNADataset(nt_features=nt_features,
                     nt_targets=nt_targets, 
                     representations=[graph_rep])
>>> dataset[0]["graph"]

{Graph(num_nodes=24, num_edges=58,
       ndata_schemes={"nt_features": Scheme(shape=(4,), dtype=torch.float32),
                      "nt_targets": Scheme(shape=(1,), dtype=torch.float32)}
        edata_schemes={"edge_type": Scheme(shape=(), dtype=torch.int64)})}
\end{lstlisting}

The example above illustrates an \texttt{RNADataset} that holds a graph representation and one input feature corresponding to a 1-hot encoding of the nucleotide identity and one binary target variable indicating whether the residue is in contact with an ion.

More generally, \texttt{rnaglib.representations.Representation} class holds the logic for converting a dataset to one of the above representations and users can easily sub-class this to create their own representations.
These classes come with their own set of attributes. Users can build several representations at the same time for the same dataset.

\clearpage

\begin{lstlisting}
>>> from rnaglib.representations import PointCloudRepresentation, VoxelRepresentation

>>> pc_rep = PointCloudRepresentation()
>>> voxel_rep = VoxelRepresentation(spacing=2)

>>> dataset.add_representation(voxel_rep)
>>> dataset.add_representation(pc_rep)
dataset[0].keys()

dict_keys(["rna_name", "rna", "path", "graph", "voxel", "point_cloud"])
\end{lstlisting}

As we can see, we now have additional keys in the returned dictionary corresponding to the data represented as voxels or point clouds. In our case, the RNA has 24 nucleotides and is approximately 12 Angstroms wide. Hence, \texttt{dataset[0][‘point\_cloud’]} is a dictionary that contains two grids in the PyTorch order :

\begin{lstlisting}
voxel_feats : torch.Size([4, 6, 5, 6])
voxel_target : torch.Size([1, 6, 5, 6])
\end{lstlisting}

While \texttt{dataset[0]["point\_cloud"]} is a dictionary that contains one list and three tensors :

\begin{lstlisting}
point_cloud_nodes : ["1a9n.Q.0", "1a9n.Q.1",... "1a9n.Q.9"]
point_cloud_coords : torch.Size([24, 3])
point_cloud_feats : torch.Size([24, 4])
point_cloud_targets : torch.Size([24, 1])

\end{lstlisting}

\subsection{Training Loop}

The last step is to feed the dataset to a loader object which takes care of batching and iterating over the dataset.
We provide custom splitting functions that take into account balancing of the target label distribution.
The loader will yield at each iteration, the batched tensors with input features and target variables. \footnote{
Here we assume that the user has implemented \texttt{model} and \texttt{criterion} in his or her framework of choice.}

\begin{lstlisting}
from torch.utils.data import DataLoader
from rnaglib.data_loading import split_dataset, Collater

train_set, valid_set, test_set = split_dataset(dataset, 
                                           split_train=0.7,
                                           split_valid=0.85)
collater = Collater(dataset=dataset)
train_loader = DataLoader(dataset=train_set, 
                          shuffle=True,
                          batch_size=2,
                          num_workers=0,
                          collate_fn=collater.collate)

# TRAINING LOOP
for batch in train_loader:
    pred = model(batch["graph"])
    loss = criterion(pred, batch["graph"]["graph_targets"])
    ...
\end{lstlisting}

\clearpage

\subsection{Visualizing Structures}

\begin{figure}
\centering
\includegraphics[width=.5\textwidth]{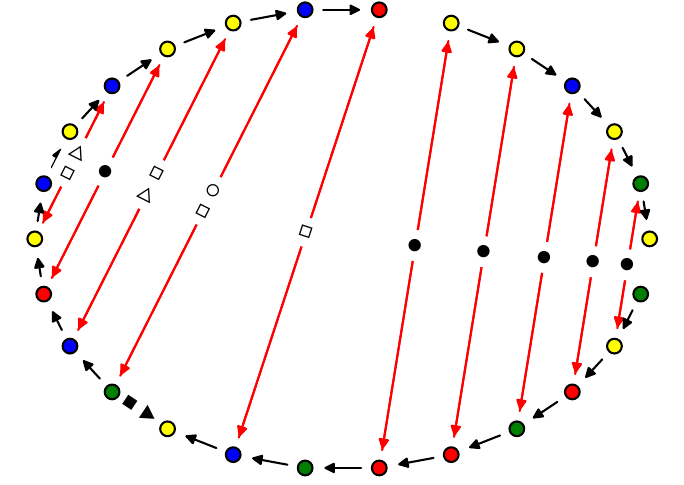}
\caption{Sample graph drawing of \texttt{PDBID: 1NLF}}
\label{fig:draw}
\end{figure}

We can return to the dataset to draw one of the graphs using the drawing package (See Fig. \ref{fig:draw} for a sample output).

\begin{lstlisting}
>>> from rnaglib.drawing import rna_draw
>>> rna_draw(dataset[358]["rna"], show=True, save="rna.pdf")
\end{lstlisting}

Note that the layout is not heavily optimized for RNA and is intended simply provided for checks and debugging.
A list of node colors can be passed with the \texttt{node\_colors} argument which can be used to visualize the model outputs on the RNA.
By default, nodes are colored according to nucleotide identity.

\section{Conclusion}

In this chapter we demonstrated the use of \rnaglib for convenient access to annotated RNA 3D structures and its use in a property prediction deep learning pipeline.
We hope this illustration will lower the barrier to entry into the emerging field of geometric deep learning on RNA 3D structures and help inform novel RNA designs and uncover new structure-function relationships from 3D structure datasets.
Finally, we point the reader again to the package documentation\footnote{\url{https://rnaglib.readthedocs.io/en/latest/}} for a complete and up to date description of functionalities, including modules for self-supervised training and evaluation, and offer an invitation to contribute new features to the tool.

\clearpage

\printbibliography

\end{document}